\documentclass[journal]{IEEEtran}
\usepackage{amsmath,amsfonts}
\usepackage{algorithmic}
\usepackage{algorithm}
\usepackage{array}
\usepackage[caption=false,font=normalsize,labelfont=sf,textfont=sf]{subfig}
\usepackage{textcomp}
\usepackage{stfloats}
\usepackage{url}
\usepackage{verbatim}
\usepackage{graphicx}
\usepackage{cite}
\begin{document}

\title{An Experimental Evaluation of Scheduled Transmission and Hardware Timestamping on the Intel E830 NIC}

\author{Takahiro Hirofuchi
\thanks{T. Hirofuchi is with Intelligent Platform Research Institute (IPRI), National Institute of Advanced Industrial Sciences and Technology (AIST), Tokyo, Japan}%
}

\maketitle

\begin{center}
\textit{Preprint version. This manuscript has not been peer reviewed.
A revised version is planned for submission to a journal or an international conference.}
\end{center}

\begin{abstract}
High-precision packet transmission is becoming increasingly important in deterministic networking applications, including 5G fronthaul and Time-Sensitive Networking (TSN).
Some recent high-speed network interface cards (NICs), including Intel E830 and NVIDIA ConnectX series,
provide hardware-assisted mechanisms for transmitting Ethernet frames accurately at specified times.
They also provide hardware timestamping for received and transmitted frames.
Although such timing capabilities are relevant to the stringent timing requirements of 5G fronthaul,
little public information is available regarding their timing accuracy and performance characteristics.

As a companion to our previous study of scheduled transmission on the NVIDIA ConnectX series, this paper presents an experimental characterization of the scheduled transmission and hardware timestamping capabilities of the Intel E830 NIC.
Using an FPGA-based measurement platform with deterministic frame generation and nanosecond-resolution timestamping, we first evaluate the precision of the receive and transmit hardware timestamps and then evaluate the accuracy of scheduled transmission.
The experimental results show that the hardware timestamping exhibits only several-nanosecond timing variation. The receive timestamp-derived frame intervals showed a peak-to-peak variation of up to 6 ns,
while the transmit timestamp comparison showed a residual range of 4.6 ns after clock-drift compensation.
Furthermore, approximately 99\% of the observed frame intervals were within ±300 ns of the target interval,
while occasional outliers of up to approximately 2 us are observed. These results indicate that the scheduled-transmission capability can support applications with timing requirements on the order of several tens of microseconds, such as 5G fronthaul, whereas its timing accuracy may be insufficient for TSN applications requiring transmission timing accuracy on the order of a few to a few tens of nanoseconds.
\end{abstract}

\begin{IEEEkeywords}
NIC, scheduled transmission, hardware timestamp, Ethernet, Intel E830.
\end{IEEEkeywords}

\section{Introduction}

Some recent high-performance network interface cards (NICs) for data centers, including the Intel E830 and NVIDIA ConnectX series, provide hardware-assisted mechanisms for scheduling the transmission time of Ethernet frames. These mechanisms appear to primarily target fronthaul communication in 5G radio access networks (RANs)\cite{ecpri}. In such networks, in-phase and quadrature (I/Q) sample data need to be transferred under stringent timing constraints\cite{ecprireq} between a distributed unit (DU), typically located at a central or edge site, and a radio unit (RU) located near the antenna.

Hardware-assisted scheduled transmission can also be applied to industrial networks\cite{P60802,8021DG,P8021DP} implementing transmission algorithms of Time-Sensitive Networking (TSN). For example, Linux's endpoint implementation of Scheduled Traffic, defined in IEEE 802.1Q-2022\cite{ieee8021q2022}, calculates frame transmission times in software and can optionally delegate the actual transmission at the specified times to NIC hardware. In our endpoint implementation of Asynchronous Traffic Shaping (ATS) also defined in IEEE 802.1Q-2022, a software state machine calculates the transmission time of each frame, and NIC hardware then transmits the frame accurately at the specified time. If data-center NICs provide sufficiently accurate scheduled transmission, they could therefore be used to implement such TSN endpoints.

However, to the best of our knowledge, vendors provide limited public information about the timing performance of the scheduled-transmission mechanisms implemented in the Intel E830 and NVIDIA ConnectX series. Although these mechanisms are expected to provide sufficient performance for applications such as 5G RAN fronthaul, their detailed timing characteristics have not been well documented. Their applicability to TSN, which imposes substantially tighter timing requirements, also remains unclear.

We are therefore conducting an experimental study of the scheduled-transmission capabilities of these high-performance NICs. As a companion to our previous study of scheduled transmission on the NVIDIA ConnectX series\cite{cx7perfarxiv}, this paper reports an experimental characterization of the corresponding functionality provided by the Intel E830.
To enable precise evaluation of transmission timing and hardware timestamping, we employ an FPGA-based Ethernet frame generation and capture platform\cite{efcc} that provides deterministic frame generation and nanosecond-resolution timestamp measurements. We first characterize the receive and transmit hardware timestamping functionalities to gain insight into the underlying timing behavior of the NIC, and then evaluate the timing accuracy of scheduled transmission on the Intel E830.

The remainder of this paper is organized as follows. Section II provides an overview of the scheduled-transmission mechanism of the Intel E830. Section III evaluates the characteristics of the receive and transmit hardware timestamps. Section IV evaluates the accuracy of scheduled transmission. Section V discusses related work. Finally, Section VI concludes the paper.

\section{Overview of Scheduled Transmission on the Intel E830 NIC}

Support for scheduled transmission on the Intel E830 was introduced in DPDK (Data Plane Development Kit), a framework for developing high-performance network applications, in version 25.07, released in July 2025. Support in the Linux kernel was introduced in version 6.18, released in November 2025. At the time of writing, a detailed datasheet for the Intel E830 is not publicly available. Although a datasheet is publicly available for its predecessor, the E810\cite{e810datasheet}, it does not describe the scheduled-transmission functionality.\footnote{Revisions 2.2 through 2.4 of the Intel E810 datasheet contained Section 10.5.4.9, titled "Tx Time (SO\_TXTIME) Overview Descriptors and Doorbells," which referred readers to Section 8.3, "Transmit Scheduling." However, Section 8.3 contained no corresponding description. In Revision 2.5, Section 10.5.4.9 was removed. Based on the available source code and other public information, scheduled transmission does not appear to have been implemented in the E810, suggesting that the corresponding description in the datasheet was likely an erratum.}
We therefore describe below our understanding of the E830 scheduled-transmission mechanism based on publicly available source code and development history, as well as our experimental observations.

When scheduled transmission is enabled, software creates, in addition to the ordinary transmit ring that contains information about frames to be transmitted, such as the memory addresses of their data, a dedicated ring for specifying their transmission times. For clarity, we refer to this ring as the Scheduled Time Queue and to each of its entries as a Scheduled Time Descriptor. In DPDK and the Linux kernel, this descriptor is referred to as a Tx time stamp descriptor ({\tt struct ice\_ts\_desc}). However, to avoid confusion with hardware transmit timestamping, which records the actual transmission time of a frame, we use the term Scheduled Time for descriptors and queues associated with scheduled transmission.
Intel NICs are configured by issuing commands to the Admin Queue, for example to create transmit queues. Creation of the Scheduled Time Queue uses Admin Queue command 0x0C35, which is not defined in the publicly available E810 datasheet.

The Scheduled Time Descriptor is 32 bits wide. A 13-bit field contains the index of the corresponding transmit descriptor in the transmit ring, while the remaining 19-bit field contains a portion of the scheduled transmission time, represented in units of 128 ns.
Consequently, only 19 bits of the scheduled transmission time, with a resolution of 128 ns, are conveyed to the hardware, corresponding to a time range of approximately 67.1 ms (128 ns x 2\textsuperscript{19} = 67,108,864 ns).

Our experiments revealed the following behaviors:
\begin{itemize}
  \item We infer that the hardware determines the transmission timing by comparing the scheduled time supplied by software with bits 7 through 25 of the nanosecond portion (the GLTSYN\_TIME\_H/L registers) of the current time maintained by the PTP Hardware Clock (PHC), where the least significant bit is numbered bit 0. This corresponds to a 19-bit time value with a resolution of 128 ns in the Scheduled Time Descriptor. Our experiments show that scheduled transmission operates correctly when the 19-bit scheduled-time value supplied to the hardware represents a time between the current PHC time and approximately 33.6 ms into the future (\(2^{25}\) ns = 33,554,432 ns), indicating that the NIC interprets such a value as a future transmission time. In contrast, when the 19-bit scheduled-time value represents a time more than approximately 33.6 ms ahead of the current PHC time, the frame is transmitted immediately. We therefore infer that such a timestamp is treated by the hardware as a time in the past.
  \item
  At the time of writing (September 2026), the latest versions of DPDK and the Linux kernel, DPDK v26.07 and Linux kernel 7.2-rc3, both contain an issue in the driver-side conversion of the scheduled transmission time, which causes transmission timing errors.
  The scheduled transmission time is maintained as a 64-bit value with 1-ns resolution. The driver first removes the seconds and higher-order components of the transmission time, effectively taking the value modulo \(10^9\) ns, and then truncates the result to the approximately 67.1-ms hardware time range by applying another modulo operation with \(2^{26}\) ns. Because \(10^9\) is not an integer multiple of \(2^{26}\), these two successive modulo operations generally produce a value different from that expected by the hardware, which corresponds to the original transmission time reduced modulo \(2^{26}\). Consequently, this mismatch in transmission-time conversion causes errors in the actual transmission timing.
  \item With both the factory-shipped firmware (NVM 1.00) and the latest firmware available at the time of writing (NVM 2.11, 0x800192cb, 1.4031.0 released on July 10, 2026), scheduled transmission is disabled when the transmission-time field in the Scheduled Time Descriptor is zero, causing the frame to be transmitted immediately. Consequently, although this occurs only very rarely, a large transmission timing error can occur when the sub-second portion of the specified transmission time is 0-127 ns.
\end{itemize}

We therefore took the following measures:
\begin{itemize}
  \item We consider the first behavior to be intentional. For scheduled transmission to operate correctly, the frame data must already be available to the NIC hardware before the specified transmission time. Software must therefore specify a transmission time sufficiently far in the future to allow the frame to reach the NIC in advance. However, specifying a transmission time more than approximately 33 ms ahead of the current time causes the problem described above. In our DPDK-based transmission program, we therefore assign each frame a transmission time approximately  100 us ahead of the PHC time expected when the frame is processed by the NIC hardware. In Linux, the ETF qdisc holds a frame until the {\tt delta} parameter before its specified transmission time and then passes it to the NIC. The value of {\tt delta} should therefore be chosen so that the frame reaches the NIC sufficiently early, while remaining below approximately 33 ms to avoid the above problem.
  \item To address the second issue, we modified the Linux driver and implemented a workaround in our DPDK-based transmission program so that the specified transmission time is mapped directly into the approximately 67.1-ms hardware time range, without first discarding the seconds and higher-order components, before being passed to the NIC hardware.
  \item To address the third issue, we also modified the Linux driver and implemented a workaround in the DPDK-based transmission program. When the specified transmission time, after being mapped into the approximately 67.1-ms hardware time range, falls within 0-127 ns, the driver writes a value of 1, rather than 0, into the transmission-time field of the Scheduled Time Descriptor. This modification can shift the transmission time specified to the hardware by up to 128 ns. However, as shown by the experimental results presented later, the actual transmission timing exhibits variations substantially larger than 128 ns. We therefore consider the impact of this modification negligible.
\end{itemize}
In this paper, scheduled transmission is evaluated using DPDK for consistency with our previous ConnectX-7 study, in which Linux did not support scheduled transmission. Hardware timestamping is evaluated using the Linux driver.

\section{Evaluation of Receive and Transmit Hardware Timestamping}

We first characterize the accuracy of the receive and transmit hardware timestamps of the Intel E830. The NIC hardware maintains the current time using a PHC, which is referenced to record the reception and transmission times of Ethernet frames with high precision. The scheduled-transmission mechanism is also believed to refer to the PHC when determining the actual time at which a frame should be transmitted. Hardware timestamping and scheduled transmission are therefore expected to share the same underlying hardware time base. Characterizing the hardware timestamping functionality thus provides insight into the timing behavior of the hardware underlying scheduled transmission.

We use the same measurement methodology as in our previous study of the NVIDIA ConnectX series. For completeness, the experimental setup is described below.
To evaluate hardware timestamping with nanosecond-level precision, we use our open-source FPGA-based Ethernet Frame Crafter and Capture (EFCC) for 10GbE. EFCC can generate arbitrary frame sequences according to predefined traffic patterns while controlling the transmission interval with a resolution of 6.4 ns. It can also record frame transmission and reception times with the same temporal resolution. In this study, EFCC is implemented on a Xilinx Alveo U250 FPGA board equipped with two QSFP28 ports. To eliminate variations in the frame interval, the Deficit Idle Count (DIC) mechanism, which dynamically adjusts the Interframe Gap (IFG) while maintaining the average IFG required by IEEE 802.3, is not used in EFCC.

The Intel E830 NIC was installed in a PCI Express slot of a host machine equipped with an Intel Xeon w5-3435X processor (16 CPU cores). The evaluated NIC was an Intel E830-XXVDA2 with firmware version NVM 2.11. To improve the reproducibility of the experimental results, Simultaneous Multithreading (SMT), Dynamic Voltage and Frequency Scaling (DVFS), and CPU idle states were disabled. PCI Express Active State Power Management (ASPM) remained disabled, which is the default setting on the system. The host ran Ubuntu 24.04 LTS with Linux kernel 6.8.0-110-generic, and DPDK v26.07 was used for packet transmission.
For the scheduled-transmission evaluation, we developed a frame transmission program using DPDK. To eliminate interference from other operating system tasks, a dedicated CPU core was isolated from the Linux scheduler at boot time and exclusively assigned to the transmission program throughout the experiments. The E830-XXVDA2 requires a PCIe Gen4 x8 interface. The host motherboard, a Supermicro X13SWA-TF, provides a PCIe Gen5 x16 slot and therefore satisfies this requirement.

According to the Intel E810 datasheet\cite{e810datasheet}, the transmit and receive hardware timestamps provided to software consist of a 32-bit field representing the time in nanoseconds and a 7-bit field representing the sub-nanosecond portion. Thus, the timestamp format has a nominal resolution of \(1/128\) ns, or 7.8125 ps. However, the 1588 timers in the E810 and its PHYs operate with an 812.5-MHz clock, corresponding to a sampling interval of approximately 1.23 ns. Both transmit and receive timestamps are sampled in the PHY.
From the publicly available driver source code, we infer that the E830 provides essentially the same timestamp interface as the E810, although the procedure for retrieving transmit timestamps differs slightly. In both DPDK v26.07 and Linux kernel 7.2-rc3, the sub-nanosecond portion of the timestamp is discarded by the driver.

Receive hardware timestamps are delivered to software through receive descriptors. When software prepares the NIC to receive frames, it creates receive descriptors in host memory and provides the NIC hardware with the physical addresses of memory buffers into which received frames are to be stored. When a frame is received, the NIC writes the frame data to the specified memory buffer and overwrites the corresponding receive descriptor with a write-back format containing completion information, including the receive hardware timestamp.

The NIC hardware provides 64 registers per port for storing transmit hardware timestamps. In the transmit context descriptor, which conveys additional control information for a frame to be transmitted, software specifies the index of the register in which the frame's transmit timestamp is to be stored. After the frame has been transmitted, software reads the specified register to obtain the transmit timestamp.

Evaluating the accuracy of NIC hardware timestamps requires particular care in conventional Ethernet, where the two endpoints generally operate with independent clocks. We therefore use different but complementary approaches for evaluating receive and transmit hardware timestamping:
\begin{itemize}
\item For receive hardware timestamping, we transmit a sequence of back-to-back frames with a strictly controlled frame interval and examine whether the intervals derived from consecutive receive hardware timestamps match the corresponding theoretical values. This approach does not directly determine the absolute timestamping error of each received frame; rather, it characterizes the consistency of the time difference between consecutive receive timestamps. Its advantage is that it avoids the need for explicit compensation for the clock offset and long-term drift between the transmitting and receiving devices.
More precisely, the actual frame interval is determined by the transmitter clock, whereas the receive timestamps are measured using the E830 time base. However, the effect of their relative clock-frequency offset over a single back-to-back frame interval is negligible for the present evaluation.
Such an evaluation is enabled by EFCC, which can generate back-to-back frames deterministically at the 10GbE line rate.
\item The same approach is difficult to apply to transmit hardware timestamping. The E830 cannot reliably provide transmit hardware timestamps for frames transmitted continuously at the line rate. One possible alternative would be to transmit frames at a lower but precisely controlled rate.
However, software-based transmission alone cannot provide sufficiently deterministic timing to establish such a reference interval. The E830 also provides scheduled transmission, which could in principle be used to generate regularly spaced transmissions. However, as shown later in this paper, the scheduled-transmission mechanism itself introduces non-negligible timing variation. It therefore cannot provide a sufficiently precise reference transmission interval for evaluating the accuracy of transmit hardware timestamping.
We therefore use an external reference for the transmit-timestamp evaluation. Frames transmitted by the E830 while recording their transmit hardware timestamps are received by EFCC through a direct DAC connection, whose propagation-delay variation is negligible for this evaluation. The E830 transmit timestamps are then compared with the corresponding receive timestamps recorded by EFCC after compensating for the relative clock offset and drift between the two devices.
\end{itemize}
These two approaches have different strengths and limitations and are therefore complementary. The receive-timestamp experiment avoids regression-based clock-drift compensation but evaluates the differences between consecutive timestamps rather than the absolute error of individual timestamps. In contrast, the transmit-timestamp experiment evaluates individual timestamps against an external reference but requires compensation for the relative clock drift between the two devices. As shown below, the fact that the two complementary approaches yield timing variations on the same order of several nanoseconds provides additional confidence in the validity of the measurements.

\subsection{Evaluation of receive hardware timestamping}

To evaluate the accuracy of the receive hardware timestamping functionality, one port of EFCC was connected to a port of the E830 NIC using a QSFP+ to 4 SFP+ breakout DAC cable. EFCC transmitted 50,000 fixed-size frames back-to-back at the 10GbE line rate. The E830 NIC recorded the receive hardware timestamp of each frame, and the frame intervals derived from these timestamps were compared with the corresponding theoretical values determined by the frame size. Receive hardware timestamps were collected using {\tt tcpdump} with hardware timestamping enabled.

For experiments using the minimum Ethernet frame size of 64 bytes, multiple UDP flows were generated by incrementing the source UDP port number for each transmitted frame. This configuration distributes incoming frames across multiple receive queues through the Receive Side Scaling (RSS) mechanism of the E830, thereby preventing frame drops caused by receive queue overflow.

\begin{table*}[t]
  \centering
  \caption{Distribution of frame intervals derived from E830 receive hardware timestamps for back-to-back transmission of 64-byte frames}
  \label{tab:rxts64b}
  \begin{tabular}{c|rrrrr}
    \hline
	  Frame interval (ns) & 69 & 70 & 71 & 72 & 73  \\
	  Frequency           & 10716  & 20441 & 12253 & 1330 & 5259 \\ \hline
	  Mean (ns) & \multicolumn{5}{c}{70.40} \\
	  Std. (ns) & \multicolumn{5}{c}{1.16} \\
    \hline
  \end{tabular}
\end{table*}

\begin{table*}[t]
  \centering
  \caption{Distribution of frame intervals derived from E830 receive hardware timestamps for back-to-back transmission of 1518-byte frames}
  \label{tab:rxts1518b}
  \begin{tabular}{c|rrrrrrr}
    \hline
	  Frame interval (ns) & 1232 & 1233 & 1234 & 1235 & 1236 & 1237 & 1238  \\
	  Frequency & 1075 & 3957 & 19580 & 3927 & 7908 & 8313 & 5239 \\\hline
	  Mean (ns) & \multicolumn{7}{c}{1235.19} \\
	  Std. (ns) & \multicolumn{7}{c}{1.62} \\
    \hline
  \end{tabular}
\end{table*}

The results for frame sizes of 64 and 1518 bytes are summarized in Tables \ref{tab:rxts64b} and \ref{tab:rxts1518b}, respectively. At the physical layer, a 64-byte Ethernet frame occupies 84 bytes on the wire, including the Preamble, Start Frame Delimiter (SFD), and Interframe Gap (IFG), and therefore requires 11 clock cycles (70.4 ns) for transmission over 10GbE. Similarly, a 1518-byte frame occupies 1538 bytes on the wire and requires 193 clock cycles (1235.2 ns). In all experiments, the mean frame interval derived from the receive hardware timestamps closely matched the corresponding theoretical value.

The difference between the maximum and minimum frame intervals was 4 ns for 64-byte frames and 6 ns for 1518-byte frames. These variations are smaller than the 12-ns range observed in our previous experiments with the NVIDIA ConnectX-7 operating in the free-running clock mode. In the E830, both receive and transmit timestamps are believed to be sampled in the PHY.
In many Ethernet NIC designs, the PHY and MAC may operate in different clock domains. The detailed clocking architecture of the E830 is not publicly documented. If such a clock-domain boundary exists in the E830 and the two domains are not phase-locked, timestamping directly in the PHY could avoid cycle-level timing variation that might otherwise arise when frame reception events are transferred to the MAC before being timestamped.

The distributions obtained from the 1518-byte experiment using a single flow and the 64-byte experiment using multiple flows exhibited nearly identical characteristics. This result indicates that the RSS mechanism does not have a measurable impact on the accuracy of receive hardware timestamping.

In contrast to the ConnectX-7, which exhibited a unimodal distribution around the theoretical frame interval with a spread of approximately 12 ns, the E830 showed no such broad distribution. This suggests that clock-cycle-level timing variation before timestamp capture is less pronounced in the E830, possibly because timestamps are sampled directly in the PHY.

\subsection{Evaluation of transmit hardware timestamping}

The E830 NIC and EFCC remained directly connected as in the previous experiment. Using the receive hardware timestamps recorded by EFCC as a reference, we evaluated the accuracy of the transmit hardware timestamping functionality of E830.

A frame transmission program was executed on the host connected to the E830 NIC. The Linux socket option for transmit hardware timestamping was enabled, and 10,000 1518-byte frames were transmitted at a constant interval under software control. After transmission, the hardware transmit timestamp of each frame was retrieved from the socket error queue. To avoid losing transmit timestamps due to insufficient socket buffer space, the socket receive buffer size (SO\_RCVBUF) was increased, as a precaution, from the default 208 KiB to 128 MiB.

To evaluate the accuracy of transmit hardware timestamping independently of the scheduled-transmission functionality, frame transmission timing was controlled solely by software. Consequently, the actual transmission times were not perfectly periodic and could exhibit sub-millisecond variations due to software scheduling. If transmit hardware timestamping operates correctly, however, it should accurately record the actual transmission time of each frame regardless of when the frame is transmitted.

Since the E830 NIC and EFCC are connected through the same breakout DAC cable, the propagation delay between them is essentially constant, except for a small variation of at most a few clock cycles caused by factors such as clock-domain crossings within the devices. Therefore, after compensating for the clock drift between the two devices, the variation in the difference between the transmit hardware timestamps reported by E830 and the receive hardware timestamps recorded by EFCC provides an estimate of the accuracy of the transmit hardware timestamping functionality.

\begin{figure}[!t]
\centering
\includegraphics[width=\columnwidth]{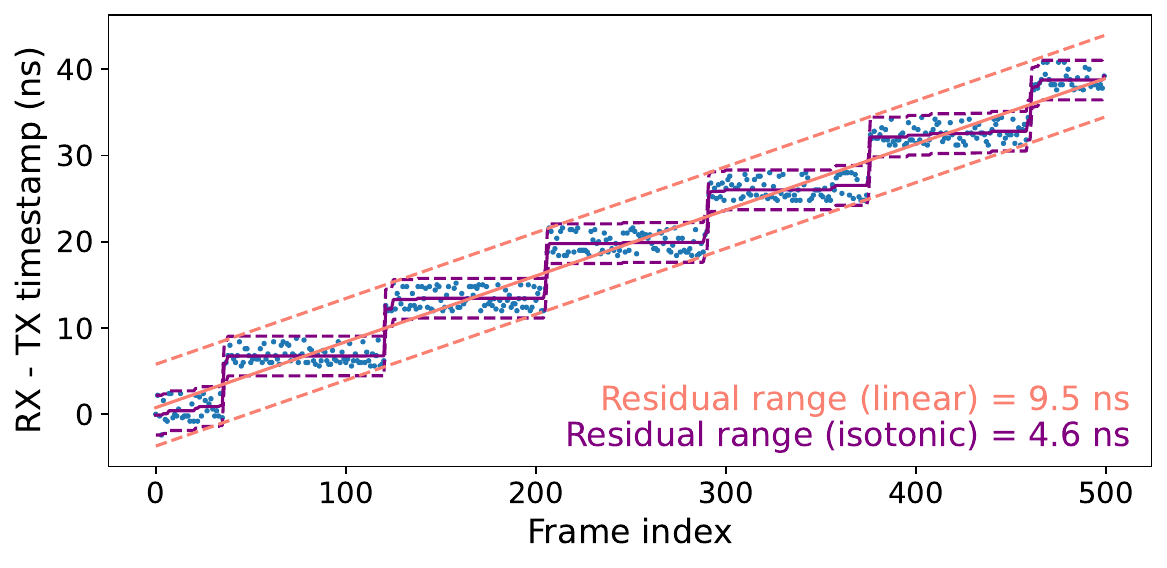}
\caption{Relative difference between the E830 transmit hardware timestamps and the reference receive hardware timestamps, normalized to the first frame. The solid lines show the linear (red) and isotonic (purple) regressions, and the corresponding dashed lines indicate their residual ranges.}
\label{fig:txts}
\end{figure}

The frame transmission interval was set to 10 us. Figure \ref{fig:txts} shows the difference between the transmit hardware timestamp reported by the E830 and the receive timestamp recorded by EFCC, normalized to the value for the first frame. For clarity, only the first 500 of the 10,000 transmitted frames are shown, although the same behavior was observed throughout the experiment.
The solid lines show the linear (red) and isotonic (purple) regression results, while the corresponding dashed lines indicate the ranges of their residuals.

The apparent propagation delay relative to that of the first frame increased approximately linearly over time. This trend was caused by the clock drift between the E830 NIC and EFCC. Specifically, the clock of the transmitting side (E830) ran slightly slower than that of the receiving side (EFCC). For example, by the 10,000th frame, which was transmitted approximately 100 ms after the first frame, the transmit-side clock lagged behind the receive-side clock by approximately 755 ns. Assuming that the oscillator frequency offset of each device remains constant under stable temperature conditions, the relative clock drift can be accurately modeled and compensated for using a linear function.

A closer inspection reveals that the apparent propagation delay exhibits a staircase-like distribution. This staircase pattern becomes more evident when the data are fitted using isotonic regression. The observed step size is approximately 6.4 ns, consistent with the temporal resolution of the EFCC receive hardware timestamp, which records frame arrival times with a resolution of one clock cycle.

The upper and lower bounds of the residuals from the linear and isotonic regressions are also shown in Figure \ref{fig:txts}.
The residual ranges of the linear and isotonic regressions were 9.5 ns and 4.6 ns, respectively. The latter indicates that, after compensating for the relative clock drift using isotonic regression, the remaining residual variation had a peak-to-peak range of 4.6 ns. This is comparable in magnitude to the 6-ns peak-to-peak variation observed in the frame intervals in the receive hardware timestamping experiment.
The similar magnitude of these variations experimentally supports the use of a common timing source for receive and transmit timestamping in the E830, consistent with the architecture described in the E810 datasheet.
The same behavior was observed when the frame transmission interval was increased to 100 us and 1 ms.

In the experiment with a 10-us transmission interval, transmit timestamps were successfully obtained for all 10,000 frames. We then modified the transmission program by removing the 10-us sleep between successive {\tt sendmsg()} calls and transmitting 10,000 frames in a busy loop. Under this condition, transmit timestamps could not be obtained for 57 frames. In all cases, the missing timestamps were caused by exhaustion of the hardware registers used to store transmit timestamps. The resulting physical-layer transmission rate, including the Preamble, SFD and IFG, was approximately 4.6 Gbps, corresponding to an average frame interval of 2,663 ns.

The transmit hardware timestamping functionality appears to be designed primarily for sporadically transmitted messages, such as PTP packets. For example, the Intel I210 GbE NIC provides only a single register for storing a transmit timestamp. In our previous experience with the I210, a timestamping rate of approximately 100 frames per second could be sustained reliably. This corresponds to a transmission interval of 10 ms, or a physical-layer data rate of approximately 1.23 Mbps for 1518-byte frames. In the E830, the number of registers available for storing transmit timestamps has increased to 64, which likely contributes to the substantially higher rate at which transmit timestamps can be obtained.

\subsection*{\bf Summary:}
The experimental results show that the hardware timestamping exhibits only several-nanosecond timing variation. The receive timestamp-derived frame intervals showed a peak-to-peak variation of up to 6 ns,
while the transmit timestamp comparison showed a residual range of 4.6 ns after clock-drift compensation.
These complementary measurements consistently indicate timing variation on the order of several nanoseconds, although the transmit-side result includes uncertainty from the comparison of independently clocked devices.
These results indicate smaller timestamp-related variation than that observed in our previous experiments with the NVIDIA ConnectX-7, as discussed in Section \ref{sec:related}.

In the E830, timestamps are believed to be captured in the PHY, and the timestamping logic is driven by an 812.5-MHz clock. These characteristics may differ from the internal timestamping design of the ConnectX-7, whose implementation is not publicly documented, and may contribute to the smaller timestamping variation observed with the E830.

\section{Evaluation of Scheduled Transmission}

Using the same experimental setup as in the previous experiment, we evaluated the timing accuracy of scheduled transmission on the E830 by analyzing the actual frame intervals measured by EFCC for frames whose transmission times were specified at constant intervals.
Using the same approach as in the transmit hardware timestamping experiment, it is possible to evaluate the accuracy of scheduled transmission by comparing the transmission times specified using the E830 clock with the frame reception times recorded by EFCC, while compensating for the relative clock offset and drift between the two devices. However, this approach requires the assumption that the relative clock-frequency offset remains stable during the measurement, for example under stable temperature conditions. Moreover, when an anomalous timing deviation is observed, it can be difficult to determine whether it originates from scheduled transmission itself or from the accuracy of the regression used for clock-drift compensation.
We therefore evaluate scheduled transmission primarily using the actual frame intervals observed by EFCC. Although this approach does not directly measure the error between each specified and actual transmission time, it avoids the need for regression-based compensation of the relative clock drift over the entire experiment and enables a more direct characterization of frame-to-frame timing variation. The remaining effect of the relative clock-frequency offset over each frame interval is sufficiently small for the approximate evaluation of scheduled-transmission accuracy presented here.

A frame transmission program was implemented using DPDK and executed on the host connected to the E830 NIC. The program transmitted 50,000 frames, each associated with a specified transmission time. The transmission time of each frame was specified such that consecutive frames were scheduled at 100-us intervals. Experiments were conducted using both 64-byte and 1518-byte frames.
For a transmission interval of 100 us, some specified transmission times are not integer multiples of 128 ns, which is the granularity at which transmission times are provided to the NIC hardware. We therefore also conducted experiments with transmission intervals of 99,968 ns and 100,096 ns, for which the specified transmission time of every frame is an integer multiple of 128 ns. Experiments were conducted using both 64-byte and 1518-byte frames.

\begin{table}[t]
  \centering
  \caption{Distribution of actual frame intervals for 64-byte frames}
  \label{tab:as64b}
  \begin{tabular}{c|rrr}
    \hline
                  & \multicolumn{3}{c}{Frame Interval (ns)} \\
                  &  99,968   & 100,000     & 100,096 \\ \hline
  Min (ns)        &  99,577.6  &   99,680.0   &   99,827.2 \\
  $P_{0.5}$ (ns)  &  99,872.0  &   99,878.4   &  100,006.4 \\
  $P_{99.5}$ (ns) & 100,057.6  &  100,160.0   &  100,185.6 \\
  Max (ns)        & 100,332.8  &  100,256.0   &  100,358.4 \\ \hline
  Max - Min (ns)  &    755.2  &     576.0       & 531.2 \\
  Mean (ns)       &  99,968.7  & 100,000.7    & 100,096.7 \\
  Std. (ns)       &  35.6     & 63.5        & 28.6 \\
    \hline
  \end{tabular}
\end{table}

\begin{table}[t]
  \centering
  \caption{Distribution of actual frame intervals for 1518-byte frames}
  \label{tab:as1518b}
  \begin{tabular}{c|rrrr}
    \hline
                  & \multicolumn{3}{c}{Frame Interval (ns)} \\
                  & 99,968   & 100,000     & 100,096 \\ \hline
	Min (ns)        &  97,977.6	&	 98,464.0	&	 98,086.4	\\
	$P_{0.5}$ (ns)  &  99,737.6	&	 99,744.0	&	 99,865.6	\\
	$P_{99.5}$ (ns) & 100,204.8	&	100,281.6	&	100,326.4	\\
	Max (ns)        & 102,073.6	&	101,632.0	&	102,060.8	\\ \hline
	Max - Min (ns)  &   4,096.0	&	  3,168.0	&	  3,974.4	\\
	Mean (ns)       & 99,968.7	&	100,000.7	&	100,096.7	\\
	Std. (ns)       & 81.2	&	99.1	&	78.0	\\
    \hline
  \end{tabular}
\end{table}

Tables \ref{tab:as64b} and \ref{tab:as1518b} summarize the measured frame intervals observed by EFCC for 64-byte and 1518-byte frames, respectively. In all configurations, the mean frame interval closely matched the intended interval between the specified transmission times, with a difference of approximately 0.7 ns. This indicates that the scheduled-transmission mechanism maintains the intended timing on average, while allowing frame-to-frame variation in the actual transmission intervals.

On the other hand, the standard deviation ranged from 28.6 to 63.5 ns for 64-byte frames and from 78.0 to 99.1 ns for 1518-byte frames, indicating that the actual transmission intervals exhibited a certain amount of jitter. In addition, the difference between the maximum and minimum observed frame intervals ranged from 531.2 to 755.2 ns for 64-byte frames and from 3,168.0 to 4,096.0 ns for 1518-byte frames. Despite this jitter, the observed timing variation was substantially smaller than that of software-based transmission, which typically exhibits sub-millisecond timing variations. The results also show that the timing variation was smaller for shorter frames.

The variation in the E830 frame intervals tended to be smaller than that observed in our previous experiments with the NVIDIA ConnectX-7. Although the ConnectX-7 also achieved the intended frame interval on average, individual frame intervals exhibited substantially larger variations. For example, in its Send Queue mode, the standard deviations were 205.8 ns for 64-byte frames and 245.2 ns for 1518-byte frames, while the differences between the maximum and minimum frame intervals were 4,718 ns and 6,995 ns, respectively. Unlike the E830, the ConnectX-7 did not exhibit a clear dependence of the timing variation on frame size.

\begin{figure}[!t]
\centering
\includegraphics[width=\columnwidth]{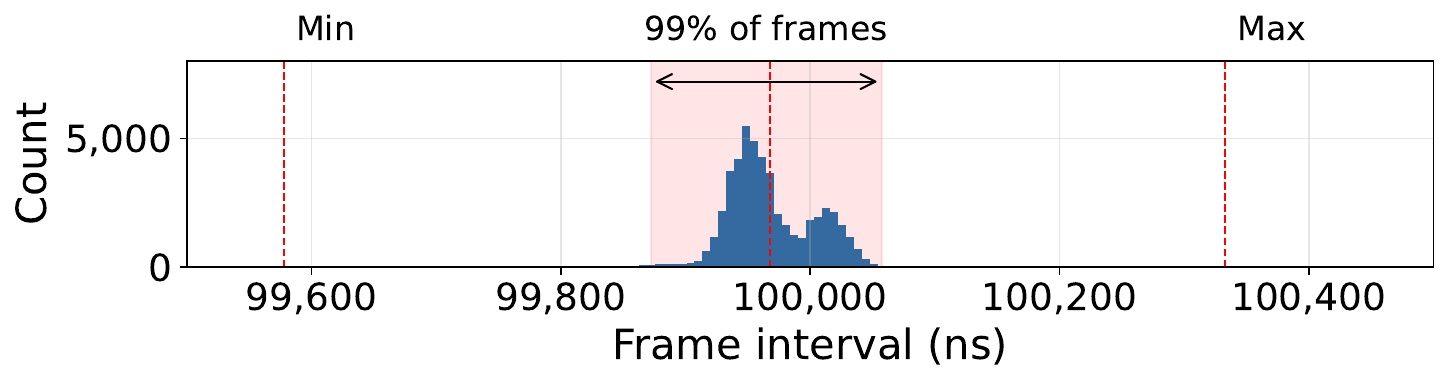}
\caption{Distribution of actual frame intervals for 64-byte frames with transmission times specified at 99,968-ns intervals}
\label{fig:asdist84int99968}
\end{figure}

\begin{figure}[!t]
\centering
\includegraphics[width=\columnwidth]{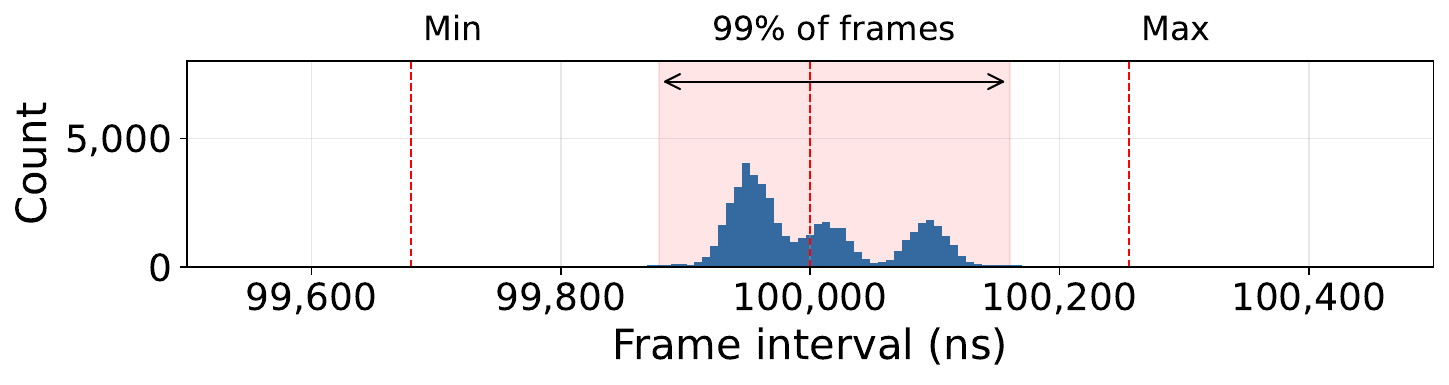}
\caption{Distribution of actual frame intervals for 64-byte frames with transmission times specified at 100,000-ns intervals}
\label{fig:asdist84int100000}
\end{figure}

\begin{figure}[!t]
\centering
\includegraphics[width=\columnwidth]{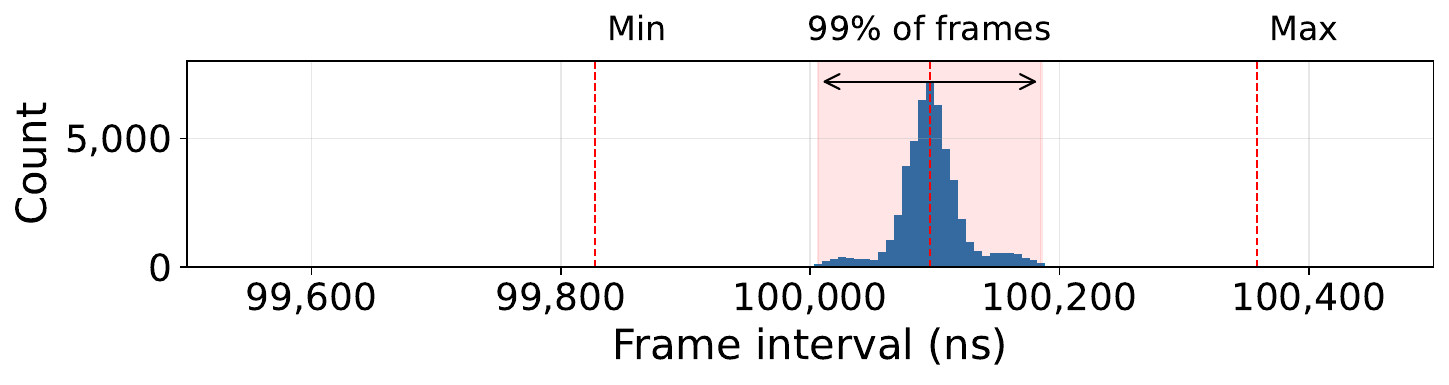}
\caption{Distribution of actual frame intervals for 64-byte frames with transmission times specified at 100,096-ns intervals}
\label{fig:asdist84int100096}
\end{figure}

\begin{figure}[!t]
\centering
\includegraphics[width=\columnwidth]{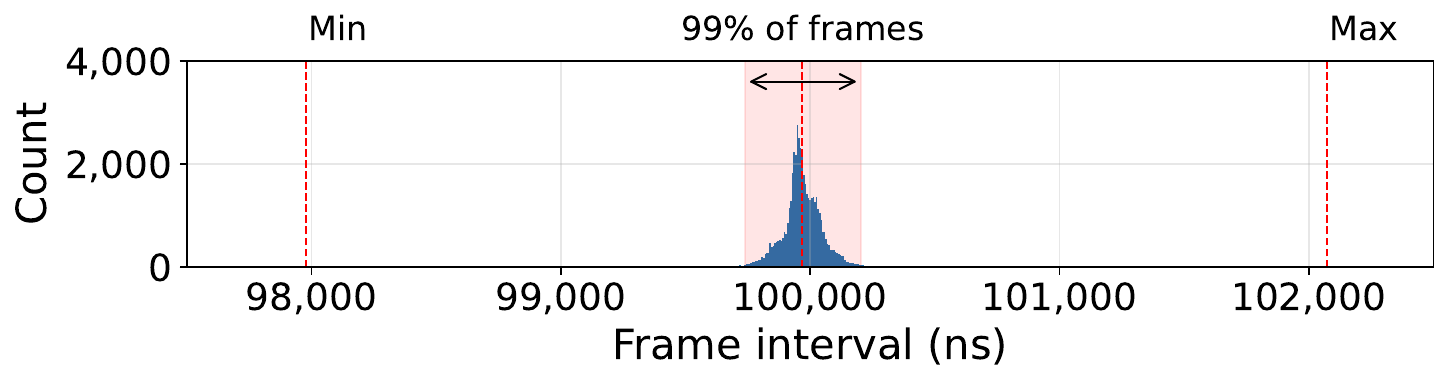}
\caption{Distribution of actual frame intervals for 1518-byte frames with transmission times specified at 99,968-ns intervals}
\label{fig:asdist1538int99968}
\end{figure}

\begin{figure}[!t]
\centering
\includegraphics[width=\columnwidth]{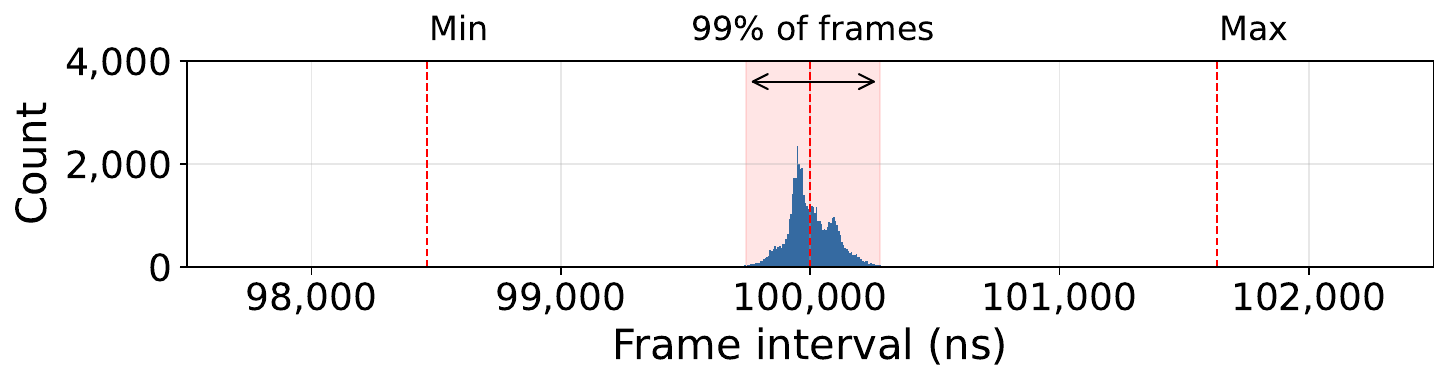}
\caption{Distribution of actual frame intervals for 1518-byte frames with transmission times specified at 100,000-ns intervals}
\label{fig:asdist1538int100000}
\end{figure}

\begin{figure}[!t]
\centering
\includegraphics[width=\columnwidth]{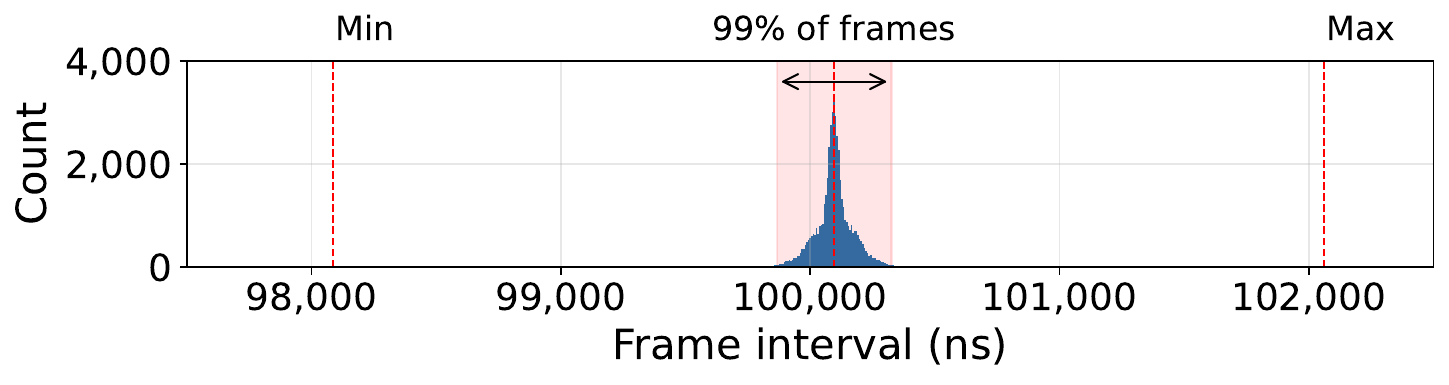}
\caption{Distribution of actual frame intervals for 1518-byte frames with transmission times specified at 100,096-ns intervals}
\label{fig:asdist1538int100096}
\end{figure}

Figures \ref{fig:asdist84int99968}--\ref{fig:asdist84int100096} show histograms of the actual frame intervals for 64-byte frames, while Figures \ref{fig:asdist1538int99968}--\ref{fig:asdist1538int100096} show those for 1518-byte frames. The histogram bin width is 6.4 ns, corresponding to the temporal resolution of EFCC.
The difference in distribution shape was more clearly observed for 64-byte frames. With a target frame interval of 99,968 ns, the distribution exhibited two distinct peaks, around 99,946 ns and 100,010 ns. The width between the 0.5th and 99.5th percentiles was 186 ns. In contrast, with a target frame interval of 100,096 ns, the distribution was approximately symmetric and unimodal, with a peak around 100,093 ns, and the width between the 0.5th and 99.5th percentiles was 179 ns.

Even when all specified transmission times were aligned to 128-ns boundaries, a certain amount of timing variation remained. Moreover, the distribution was bimodal only for the 99,968-ns interval, whereas a nearly symmetric unimodal distribution was observed for the 100,096-ns interval. This difference suggests that some aspect of the internal scheduled-transmission implementation may influence the distribution of transmission timing variation. Nevertheless, in both cases, the mean frame interval closely matched the intended value, indicating that the mechanism is designed to maintain the target interval with high accuracy on average.

A similar tendency was also observed for 1518-byte frames, although it was less clearly visible in the histograms. These results indicate that aligning the specified transmission times to 128-ns boundaries can help reduce timing variation. However, a non-negligible amount of variation remains even under this condition, suggesting that some component within the NIC hardware may introduce transmission-time variation on a comparable time scale.

\begin{figure}[!t]
\centering
\includegraphics[width=\columnwidth]{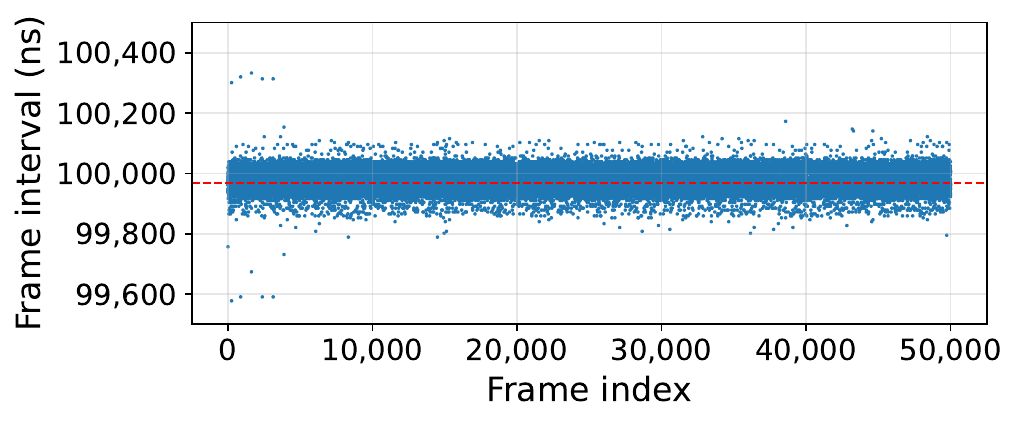}
\caption{Actual frame intervals versus frame index for 64-byte frames with transmission times specified at 99,968-ns intervals}
\label{fig:astimeseries84int99968}
\end{figure}

\begin{figure}[!t]
\centering
\includegraphics[width=\columnwidth]{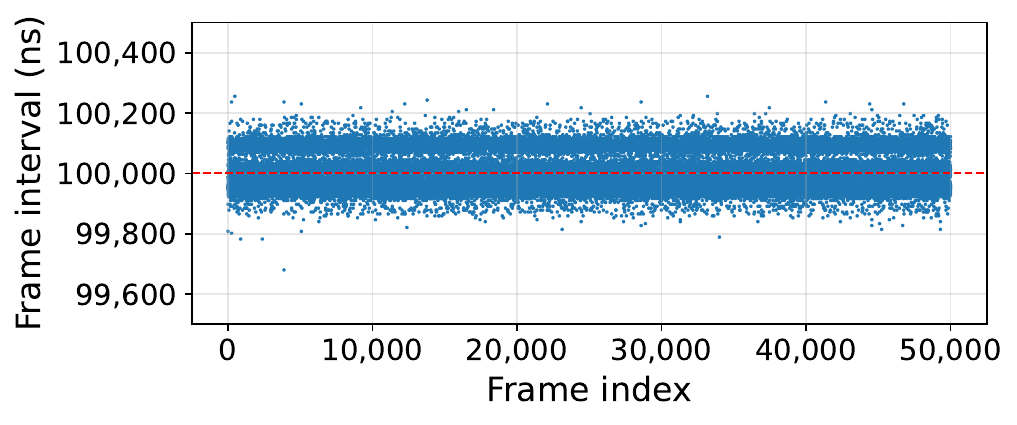}
\caption{Actual frame intervals versus frame index for 64-byte frames with transmission times specified at 100,000-ns intervals}
\label{fig:astimeseries84int100000}
\end{figure}

\begin{figure}[!t]
\centering
\includegraphics[width=\columnwidth]{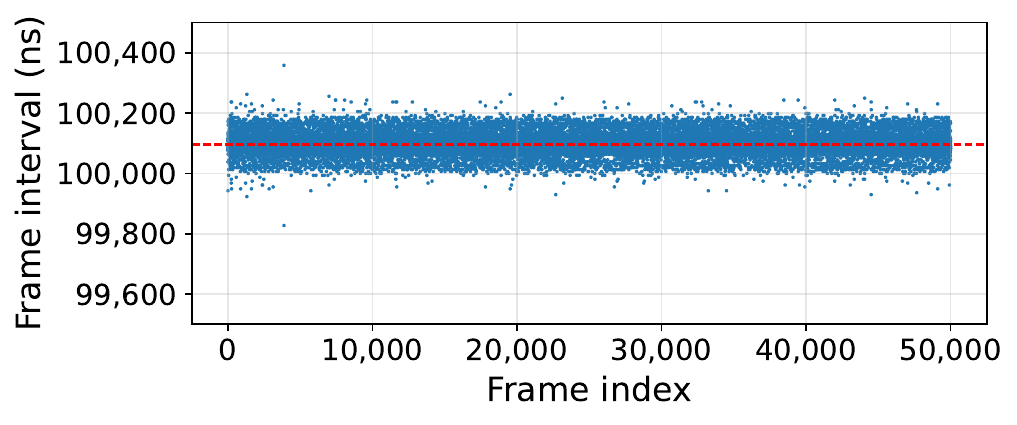}
\caption{Actual frame intervals versus frame index for 64-byte frames with transmission times specified at 100,096-ns intervals}
\label{fig:astimeseries84int100096}
\end{figure}

\begin{figure}[!t]
\centering
\includegraphics[width=\columnwidth]{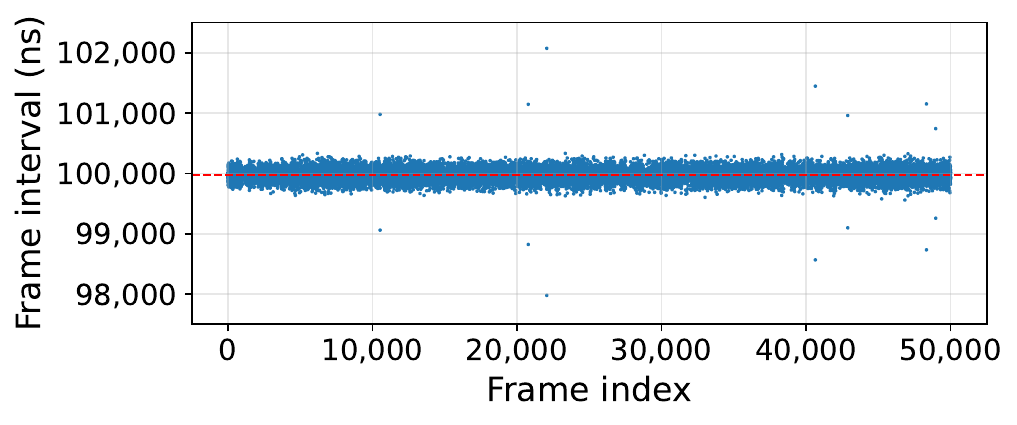}
\caption{Actual frame intervals versus frame index for 1518-byte frames with transmission times specified at 99,968-ns intervals}
\label{fig:astimeseries1538int99968}
\end{figure}

\begin{figure}[!t]
\centering
\includegraphics[width=\columnwidth]{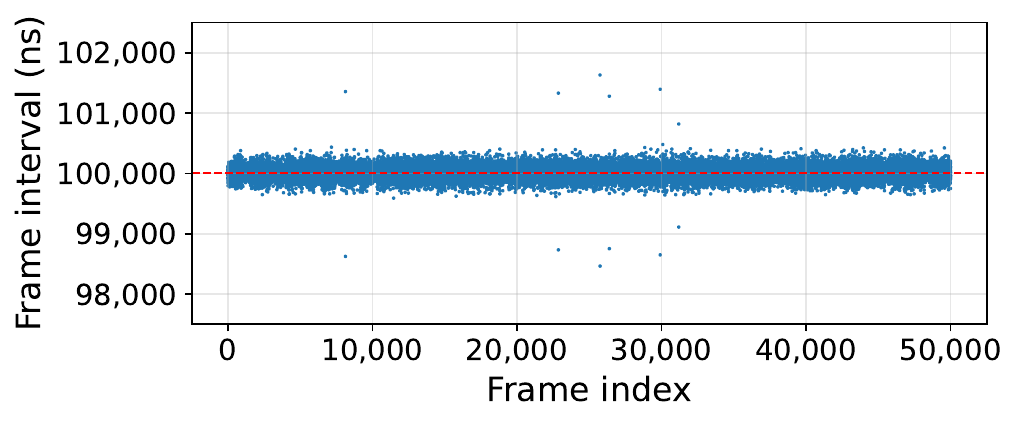}
\caption{Actual frame intervals versus frame index for 1518-byte frames with transmission times specified at 100,000-ns intervals}
\label{fig:astimeseries1538int100000}
\end{figure}

\begin{figure}[!t]
\centering
\includegraphics[width=\columnwidth]{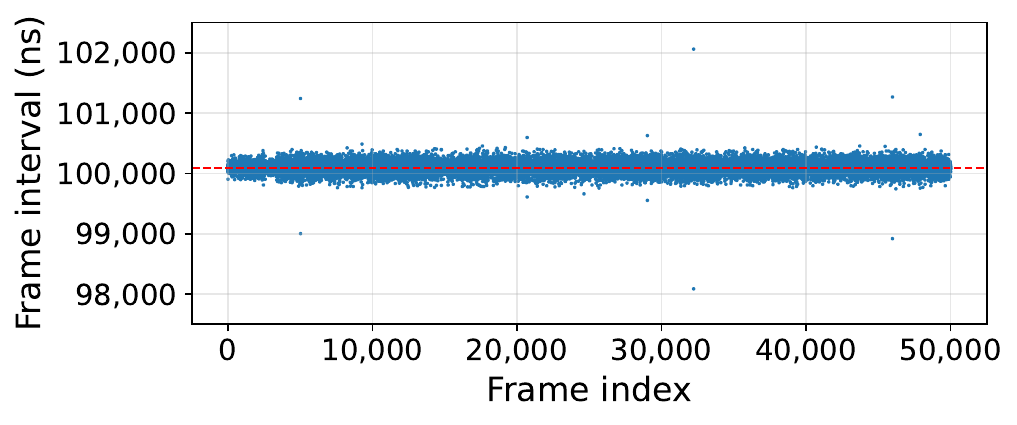}
\caption{Actual frame intervals versus frame index for 1518-byte frames with transmission times specified at 100,096-ns intervals}
\label{fig:astimeseries1538int100096}
\end{figure}

Figures \ref{fig:astimeseries84int99968}--\ref{fig:astimeseries1538int100096} show the actual frame interval of each transmitted frame. Rare outliers occurred sporadically throughout the experiments. The deviations from the target interval reached approximately 400 ns for 64-byte frames and approximately 2 us for 1518-byte frames. In our previous experiments with the ConnectX-7, in addition to sporadic outliers, the frame intervals sometimes exhibited an apparent periodic variation over time. No such periodic behavior was observed with the E830.

\subsection*{\bf Summary:}

Overall, the experimental results demonstrate that approximately 99\% of the observed frame intervals were within \(\pm\)300 ns of the target interval. Although occasional outliers with deviations of up to approximately 2 us were observed, they occurred only rarely. This level of timing precision is difficult to achieve using software-based transmission alone and is expected to be sufficient for applications with timing requirements on the order of several tens of microseconds, such as 5G fronthaul using eCPRI.
In contrast, the achieved accuracy may be insufficient for TSN applications requiring timing accuracy on the order of several to several tens of nanoseconds.
Although the exact requirement depends on the application and network configuration, the transmission-time error should be kept sufficiently smaller than the minimum frame interval (e.g. 70.4 ns at 10GbE with the constant IFG) to avoid effects such as frame reordering and increased deviation from the worst-case delay bound.

Because the transmission time is specified to the NIC hardware with a granularity of 128 ns, transmission times that are not aligned to 128-ns boundaries can introduce timing variation due to this quantization. However, even when all specified transmission times were aligned to 128-ns boundaries, timing variation of a comparable or greater magnitude remained. This indicates that additional timing variation is introduced by some internal mechanism of the NIC hardware.
The 128-ns granularity for specifying transmission times may have been chosen on the assumption that the NIC hardware itself introduces timing variation of a comparable or greater magnitude.

The measured receive and transmit timestamping variations were only on the order of several nanoseconds. The substantially smaller timestamping variation indicates that the variation observed in scheduled transmission is unlikely to originate from the accuracy of the common timekeeping logic within the NIC. In this evaluation, we used the receive timestamps recorded by EFCC to make the measurement independent of the E830 timestamping mechanism. We also repeated the analysis using the E830 transmit hardware timestamps and obtained similar results. Since the E830 transmit timestamp is believed to be captured in the PHY, this result suggests that the dominant source of the scheduled-transmission variation occurs upstream of the PHY timestamping point, before the transmit timestamp is captured.

As described above, all of the experiments presented in this section were conducted after applying our own fixes to the identified E830 driver issues. Without correcting the mismatch in the modulo operations used to convert the scheduled transmission time, our experiments produced frame intervals as large as 6.7 ms. In addition, if we did not apply the workaround for the case in which scheduled transmission is ignored when the transmission-time field of the Scheduled Time Descriptor is zero, this condition occasionally occurred during the transmission of 50,000 frames. In such runs, the minimum observed frame interval became back-to-back for the frame size because a frame was transmitted immediately rather than at its scheduled time.
Both issues should be corrected in the upstream driver source code. For the latter issue, however, a firmware-level fix would likely be the more fundamental solution, because the current driver workaround merely avoids specifying a zero value in the transmission-time field.

In addition, we found that the E830 can correctly interpret scheduled transmission times only when they are within approximately 33 ms of the current time. Consequently, software must ensure that frames whose specified transmission times are more than approximately 33 ms in the future are not submitted to the NIC hardware. For example, the Linux ETF qdisc holds frames in software until shortly before their specified transmission times, while our DPDK transmission program carefully selects the transmission time assigned to each frame so that it falls within the supported future-time window when the frame is submitted to the NIC.

\section{Related Work}
\label{sec:related}

In our companion study\cite{cx7perfarxiv}, we experimentally characterized the receive and transmit hardware timestamping and scheduled-transmission capabilities of the NVIDIA ConnectX series.
For receive hardware timestamping, the frame intervals derived from the ConnectX-7 receive timestamps exhibited peak-to-peak variations of 12 ns in the free-running clock mode and 16 ns in the real-time clock mode. In comparison, the corresponding variation with the E830 was up to 6 ns. For transmit hardware timestamping, after compensating for the relative clock drift between the NIC and EFCC, the ConnectX-7 exhibited an isotonic-regression residual range of 14.4 ns. The corresponding transmit-side variation observed with the E830 was 4.6 ns. These results indicate that the E830 generally exhibits smaller timestamp-related timing variation than the ConnectX-7.
For scheduled transmission, approximately 99\% of the observed frame intervals with the ConnectX-7 were within \(\pm\)900 ns of the target interval, with rare outliers reaching approximately 5 us. In comparison, with the E830, approximately 99\% of the observed frame intervals were within \(\pm\)300 ns of the target interval, with similarly rare outliers reaching approximately 2 us. These results indicate that the E830 exhibits smaller frame-to-frame timing variation than the ConnectX-7, while the frequency of large outliers is comparable.

Similar to the Intel E830 and NVIDIA ConnectX series, several commercial Ethernet controllers provide hardware-assisted transmission scheduling capabilities for deterministic networking applications\cite{P60802,8021DG,P8021DP}. Intel Ethernet controllers such as the I210 and I225 support LaunchTime, which allows software to specify the transmission time of each frame. This functionality is intended to support software implementations of the Scheduled Traffic mechanism (commonly referred to as Time-Aware Shaping, TAS) defined in IEEE 802.1Q-2022. STMicroelectronics also provides a similar functionality called Time-Based Scheduling (TBS) in some of their Ethernet controller products. However, these controllers are limited to link speeds of up to 1 or 2.5 GbE, making them unsuitable for applications requiring 10GbE or higher, such as 5G fronthaul and low-latency user-plane processing nearby a 5G core network\cite{mec003,mec002}.

The Intel I210 and I225 are widely used in off-the-shelf GbE NICs. In particular, the I210 has been used in several TSN studies because its programming interface is publicly documented\cite{i210datasheet}, and a number of studies have reported performance evaluations of its LaunchTime functionality\cite{method2022,oge2020,dac2024}.
In the Intel I210, transmission times are specified to the NIC hardware with a granularity of 32 ns, and only times within approximately \(\pm\)500 ms of the current time can be represented. Although the E830 uses a coarser 128-ns granularity and a narrower range of approximately \(\pm\)33 ms, its scheduled-transmission mechanism appears to follow a design concept similar to that of the earlier I210.
Only a limited portion of the transmission time is passed to the hardware.

In our previous work\cite{access2025hirofuchi}, we developed an ATS endpoint based on the Intel I210 and conducted a more rigorous evaluation of its transmission timing accuracy, particularly under high-rate transmission conditions, to determine whether it could satisfy the timing requirements of ATS. We confirmed that the I210 is capable of transmitting frames at specified transmission times with an accuracy of approximately 48 ns under suitable conditions.

However, we also found that its timing accuracy depends on both the PCIe implementation of a transmitting host and frame size. In one host machine, although its PCIe interface satisfied the requirements of the I210, for example, the transmission timing accuracy for 1518-byte frames degraded rapidly when the frame interval became shorter than approximately 18 us. As the transmission interval decreased, the timing error increased to nearly 10 us. The NIC was unable to sustain the target transmission rate, resulting in frame drops.

According to the Intel I210 datasheet\cite{i210datasheet}, enabling LaunchTime changes the way the NIC fetches frame data from host memory. The datasheet also states that the controller provides a 24-KB on-chip transmit buffer.
Since the on-chip buffer is shared among multiple hardware transmit queues, using the entire 24 KB buffer to prefetch frames for a single hardware queue would prevent the NIC from transmitting frames from other hardware queues when they become eligible for transmission. Such behavior would violate the scheduling semantics required by IEEE 802.1Q, in which the interactions among different traffic classes are strictly specified.
A plausible explanation is therefore that, when LaunchTime is enabled, the NIC limits the amount of frame data prefetched from host memory for each hardware queue. The transmission process becomes more sensitive to PCIe access latency, which resulted in time errors with a particular PCIe slot.

In contrast, our experiments evaluated the scheduled transmission of the E830 and ConnectX-7 using a transmission interval of 100 us, which is substantially longer than those used in our previous evaluation of the Intel I210. The transmission timing deviations of up to approximately 2 us and 5 us observed with the E830 and ConnectX-7, respectively, are unlikely to originate from delays in fetching frame data from host memory over the PCIe bus. This suggests that the dominant source of these deviations lies within the NIC after the frame data have been fetched from host memory.

\section{Conclusion}

This paper presented an experimental characterization of the hardware timestamping and scheduled transmission capabilities of the Intel E830 NIC using an FPGA-based measurement platform with deterministic frame generation and nanosecond-resolution timestamping.
The experimental results show that the hardware timestamping exhibits only several-nanosecond timing variation. The receive timestamp-derived frame intervals showed a peak-to-peak variation of up to 6 ns,
while the transmit timestamp comparison showed a residual range of 4.6 ns after clock-drift compensation.
In addition, approximately 99\% of the observed frame intervals were within ±300 ns of the target interval, although occasional outliers of up to approximately 2 us were observed.

The substantially smaller timestamping variation suggests that the dominant source of the observed timing variation is unlikely to be the common timekeeping logic within the NIC. Although the 128-ns transmission-time granularity can contribute quantization error, substantial variation remained even when the specified transmission times were aligned to 128-ns boundaries, suggesting that additional variation arises in the internal transmission-timing control mechanism.

These results indicate that the scheduled transmission is well suited for applications with latency requirements on the order of several tens of microseconds, such as 5G fronthaul. However, its timing accuracy remains insufficient for highly deterministic TSN applications that require transmission timing accuracy on the order of several to several tens of nanoseconds. As future work, we plan to investigate whether the scheduled transmission can be applied to TSN by extending deterministic network models to incorporate bounded transmission timing errors. For example, our ATS endpoint mechanism has a configurable safety margin for transmission timing errors while remaining fully compliant with the ATS state machine defined in the IEEE standard. Although further investigation is required, this mechanism is expected to be applicable to the timing errors of the scheduled transmission.

\section*{Acknowledgments}
This paper is based on results obtained from the project, "Research and Development Project of the Enhanced infrastructures for Post-5G Information and Communication Systems" (JPNP20017), commissioned by the New Energy and Industrial Technology Development Organization (NEDO).

\bibliographystyle{IEEEtran}
\bibliography{myrefs}

\end{document}